\documentclass[numberedappendix]{emulateapj}

\newcommand{\hi}{\mbox{\rm \ion{H}{1}}}
\newcommand{\hii}{\mbox{\rm \ion{H}{2}}}

\newcommand{\permpccu}{\,{\rm Mpc^{-3}}}

\def\h2{H$_2$}

\def\c2{cm$^{-2}$}
\def\2e{$^{2}$\ }
\def\2{$^{2}$}
\def\3e{$^{3}$\ }
\def\3{$^{3}$}

\def\mone{$^{-1}$}

\def\hper{\ifmmode \rlap.^{h} \else $\rlap{.}^h$\fi}
\def\mper{\ifmmode \rlap.^{m} \else $\rlap{.}^m$\fi}
\def\sper{\ifmmode \rlap.^{s} \else $\rlap{.}^s$ \fi}
\def\degper{\ifmmode \rlap.^{\circ} \else $\rlap{.}^{\circ} $\fi}
\def\arcmper{\ifmmode \rlap.{' } \else $\rlap{.}' $\fi}
\def\arcsper{\ifmmode \rlap.{'' } \else $\rlap{.}'' $\fi}

\newcommand\subm{$_{\mathrm {M}}$}

\def\rh2{$\rho_{H_2}$}
\def\mrh2{\rho_{H_2}}
\def\rhi{$\rho_{HI}\ $}
\def\mrhi{\rho_{HI}}
\def\rext{$\rho_{ext}\ $}
\def\mrext{\rho_{ext}}

\shorttitle{Gas Consumption History}
\shortauthors{Bauermeister, Blitz, \& Ma}
\begin{document}

\title{The Gas Consumption History to Redshift 4}

\author{Amber Bauermeister, Leo Blitz and Chung-Pei Ma}
\affil{Department of Astronomy, University of California at Berkeley,
\\601 Campbell Hall MC 3411, CA 94720}
\email{amberb@berkeley.edu; blitz@berkeley.edu;
cpma@berkeley.edu}


\begin{abstract} 

Using the observations of the star formation rate and \hi\ densities to
$z \sim 4$, with measurements of the Molecular Gas Depletion Rate
(MGDR) and local density of H$_2$ at $z = 0$, we derive the history of
the gas consumption by star formation to $z \sim 4$.  We find that closed-box
models in which H$_2$ is not replenished by \hi\ require improbably large 
increases in $\rho(H_2)$ and a decrease in the MGDR with lookback time 
that is inconsistent with observations. Allowing the H$_2$
used in star formation to be replenished by \hi\ 
does not alleviate the problem because observations show that there is
very little evolution of $\mrhi(z)$ from $z = 0$ to $z = 4$.  We show that
to be consistent with observational constraints, star formation on cosmic
timescales must be fueled by intergalactic ionized gas, which may come from
either accretion of gas through cold (but ionized) flows, or from ionized
gas associated with accretion of dark matter halos.  We constrain
the rate at which the extraglactic ionized gas must be converted
into \hi\ and ultimately into H$_2$. The ionized gas inflow rate roughly traces
the SFRD: about 1 -- 2 $\times 10^8$
M$_\sun$ Gyr$^{-1}$ Mpc$^{-3}$ from $z \simeq 1 - 4$, decreasing by about
an order of magnitude from $z=1$ to $z=0$ with
details depending largely on MGDR(t). 
All models considered require the
volume averaged density of \rh2 to increase by a factor of 1.5 -- 10 to $z
\sim 1.5$ over the currently measured value.  Because the molecular gas
must reside in galaxies, it implies that galaxies at high $z$ must, on
average, be more molecule rich than they are at the present epoch, which
  is consistent with observations. These quantitative results, derived solely from
  observations, agree well with cosmological simulations.
\end{abstract}

\section{Introduction} 
\label{intro}


The time variation of the mean star formation rate in galaxies is by now
well established \citep{Madau, Steidel, Hippelein, HB2006}.
The star formation
rate is either flat or slowly rising with time, reaching a maximum near
$z \sim$ 1 -- 2, and then declines precipitously down to the
current epoch.  This change in the star formation rate must be closely
coupled to both the inventory of gas available for star formation, and
the way in which this gas is channeled into galaxies.  

Stars condense only from molecular gas at the current epoch and at all
epochs in the past.  This statement derives from both observational and
theoretical considerations.  Observationally, the youngest stars are always
found to be associated with their nascent molecular material both in the
local Universe and at high $z$ (e.g.  \citealt{blaauw1964, hr1972, swe1973,
  omont1996, carilli2002}).  The interstellar gas in star forming regions
is almost completely molecular, representing a stable phase of the ISM with
little atomic content \citep{burton1978}.  Theoretically, there is general
consensus that the initiation of star formation requires the nascent gas to
become Jeans unstable, probably mediated by magnetic fields
\citep{sal1987}.  In star forming regions, $T$ is typically 10 - 20 K, but
in any event cannot be less than 2.7 K.  In order to get a solar mass star
at 10 K, the Jeans instabiliy criterion would require a density of $\rho_J
> (kT/{\mu}{m_H}G)^3 (\pi^5/36 M_J^2) \sim 10^6 {\rm cm}^{-3}$ if a
molecular core forms a star at 100\% efficiency.  The density must be
higher if the efficiency is lower as many observations now suggest
(e.g. \citealt{man1998, all2007, myers2008}).  In order to reach
these temperatures and densities, the gas must be fully molecular in
  order 
to achieve the necessary cooling.
We would therefore expect that, in a broad sense, the gas consumption rate
is closely tied to the star formation rate.

Unfortunately, there are few constraints on the gas from
observations because little is known about the distribution of neutral gas
at high $z$.  There are very few detections of atomic gas in emission at $z
\ga$ 0.1, and what little we know about the atomic gas comes from
Lyman-alpha lines seen in absorption toward quasars and radio-loud AGN
(e.g. \citealt{PW2009,wolfe2005,zp2006}).  Molecular line observations at
high-$z$ have been largely limited to the brightest objects, though some
recent observations at $z\sim 2$ have begun to probe lower luminosity
systems \citep{fs2009,Daddi2009,t2010}.

Gas consumption by star formation in galaxies has been investigated
previously via observations of the gas depletion time, $\tau_{dep} = M_{gas} / SFR$. This 
represents the amount of time it will take the galaxy to completely
exhaust its gas supply at the current star formation rate. Studies of individual local 
disk galaxies find depletion times on the order of a few Gyr, much shorter than
the Hubble time (e.g. \citealt{Larson1980,k94}). This is the gas depletion problem:
without some form of gas replenishment, star formation in disk galaxies should
be coming to an abrupt end. One proposed solution is stellar recycling, which
\cite{k94} find can extend the gas depletion times in many local disk galaxies 
by a factor of 1.5 to 4. Gas accretion has long been
suggested as a solution to the gas depletion problem as well, and observational
support for inflow is accumulating. \cite{Sancisi2008} review the observational evidence
for gas accretion such as galaxy interactions and minor mergers, high velocity
clouds (HVCs), extra-planar gas and warped or lop-sided HI disks. These 
observations yield an estimate for the 'visible' gas accretion rate onto a typical
disk galaxy of $\sim0.2$ M$_{\odot}$ yr$^{-1}$, which falls short of the typical
star formation rate of $\sim1$ M$_{\odot}$ yr$^{-1}$. These studies focus on recycling 
and inflow in local disk galaxies, but one may ask how does gas consumption 
evolve with redshift?

The time evolution of gas in disk galaxies has been studied on large scales using 
observations of damped Ly$\alpha$ systems (DLAs) to infer the cosmic mass
density of HI as a function of redshift. The cosmic mass density is the mass density
averaged over a large volume so as to be representative of a typical Mpc$^{-3}$
of the universe. \cite{Lanzetta1995} and \cite{PF1995} build simple models
of gas consumption focusing on the chemical evolution of the gas using the
cosmic mass density of HI and observed metallicities as inputs. Their models
predict the star formation rate density (SFRD) of the universe as a function of redshift
for their chosen inflow and outflow parameters. However, 
the SFRD is now becoming increasingly well constrained by observations and we
can use SFRD$(z)$ as an input to our models in order to place constraints
on the gas inflow rates.

Finally, this problem has been approached using numerical simulations to 
explain observations of the SFRD and galaxy properties. 
The shape of the SFRD has been investigated using cosmological, 
hydrodynamical simulations that include prescriptions 
for star formation and feedback (e.g. \citealt{HS2003,Schaye2010}).
These studies suggest that the shape of the SFRD at high redshifts is
determined by the buildup of dark matter halos and the gas brought in with them, 
and the decline of the SFRD at low
redshifts is due to lower cooling rates in the halo gas, gas exhaustion and stellar
and black hole feedback. The results of simulations have also been used to build
simple models in order to better understand gas accretion to fuel star formation in 
galaxies.  \cite{Bouche2009} 
build a simple model of gas consumption starting with simulated halo growth histories 
with different prescriptions for gas accretion and star formation. The predictions 
from the different prescriptions are compared to the observed SFR-Mass 
and Tully-Fisher relations for galaxies from $z\sim2$ to $z=0$. The authors find
that the prescription for gas accretion modeled on cold flows with halo mass
cutoffs agrees best with observations.

In this paper, we look at the issue in reverse.  We build simple models of gas
consumption based solely on observations in order to understand the roles
of the different phases of gas in star formation in galaxies. Using observations at $z$
= 0, we examine which quantitative conclusions can be extrapolated to high
$z$, and using observations of the star formation history, we make several
inferences about how the gas consumption into stars must have proceeded.
We find that the relationship of the inventory of gas to that of
the stars is not straightforward: the observations imply that that all phases 
of the interstellar and intergalactic medium must be taken into account in 
order to understand how the gas forms stars. Building a model that includes 
all the gas phases, we make predictions about gas densities and their
variation at intermediate- and high-$z$, and how this gas must have been
accreted into galaxies.

The paper is organized as follows. \S\ref{obs} describes the observations
of the SFRD, MGDR, \rh2 and \rhi that we use as inputs to our models.
In \S\ref{models}, we build three models to fit the observations: the restricted
closed box model, the general closed box model and the open box model. 
In \S\ref{discussion}, we discuss potential changes to our star formation 
prescription at high redshift and examine the predictions of the open box model.
Throughout this paper, we adopt a standard $\Lambda$CDM cosmology with (h,
$\Omega_M$, $\Omega_{\Lambda}$) = (0.7, 0.3, 0.7).  All of the densities are
in comoving units.




\section{Observations}
\label{obs}

\subsection{SFRD}
\label{SFRDobs}

\begin{figure}[ht]
\includegraphics[scale=0.5]{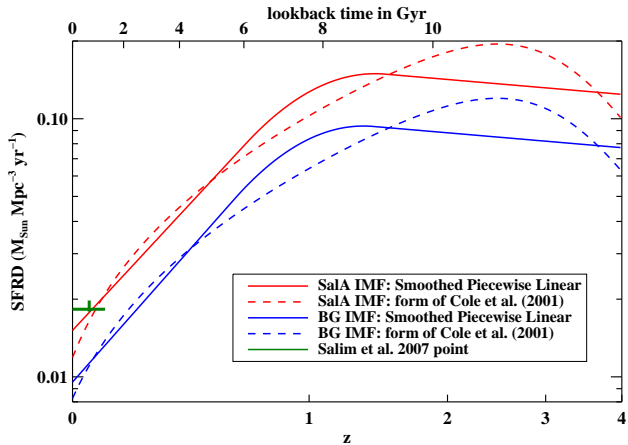}
\caption{Comparison of forms of the SFRD.  The
  solid curves show our smoothed version of the piecewise linear fits from
  \cite{HB2006}; the dashed curves show the fits to the \cite{Cole2001}
  form.  Two IMFs are plotted: the modified Salpeter A IMF (SalA IMF; red) and
  the IMF from \cite{BG2003} (BG IMF; blue).  The green symbol denotes the
  SFRD at $z = 0$ measured by \cite{Salim}.}
\label{SFRfig}
\end{figure}

The observed (comoving) star formation rate density, SFRD, was estimated by
\cite{Madau} over a large range of $z$ and re-examined by several other
investigators \citep{Steidel, Hippelein, HB2006}.  We will focus on the
results of \cite{HB2006}, a compilation of SFRD measurements at different
wavelengths. The measurements are converted to a common cosmology, SFRD
calibration, and dust obscuration correction; the data are then fit to a
piecewise linear form in $\log(1+z)$ vs. $\log(SFRD)$ space as well as the
parameterization from \cite{Cole2001}.  \cite{HB2006} find that changing
the assumed IMF corresponds to a simple change in the overall amplitude of
the SFRD, so each fit is done using two extreme IMF forms in order to
provide bounds on the actual form. These two extreme forms for the IMF are
a modified Salpeter A IMF and the form of \cite{BG2003}.  Fig.~\ref{SFRfig}
plots the fits for the two IMFs in red and blue. We have smoothed the original
piecewise linear fits from \cite{HB2006} for our models (solid lines). The form
of the smoothed piecewise linear fit is given in Appendix A. 

At $z$ = 0, the SFRD fits from \cite{HB2006} predict a local SFRD of $(0.8 -
1.5) \times 10^{-2}$ M$_\sun\permpccu$ yr$^{-1}$.  In an extensive study of
galaxies in the local universe (to $z$ = 0.1), \cite{Salim} find that the
SFRD is $1.828^{+0.148}_{-0.039} \times 10^{-2}\ h_{70}$ M$_\sun\permpccu$
yr$^{-1}$ from UV observations (using the \citealt{Chabrier} IMF),
a value they argue is the most accurate
determination of this number to date. This value (shown as a data point in
Fig.~1) is indeed close to those found from other recent investigations
\citep{Houck2007, Hanish2006} and is also in reasonable agreement with the
forms of \cite{HB2006}.

The agreement between the locally determined SFRD and extrapolation from
studies at higher $z$ \citep{HB2006, Steidel, Hippelein} is encouraging and
provides some confidence that the star formation rates and their time
variation are being measured reliably.  While somewhat different
functional forms for the decline of the SFRD with time have been proposed
in the literature, e.g., log(SFRD) has been found to be linear in $z$
\citep{Steidel}, in $t$ \citep{Hippelein}, and in $\log(1 + z)$
\citep{HB2006}, these differences have only a small quantitative effect in
what follows.

\subsection {MGDR}
\label{mgdrsec}
\subsubsection {Measurements at z = 0}

The star formation efficiency, SFE, is often defined as the star formation
rate per comoving volume divided by the mass per comoving volume of gas;
its units are yr{\mone} (\citealt{Leroy2008} and references therein).
Defined in this way, the SFE is not properly an efficiency, but a rate, and
we drop this unfortunate usage even though it has become firmly entrenched
in the observational literature.  Since we are interested in SFE\subm, the
star formation rate density divided by the density of {\it molecular} gas,
$\mrh2$, we introduce the molecular gas depletion rate, MGDR, to replace
the usage of SFE\subm. We define MGDR as SFRD divided by the density of
molecular hydrogen, \rh2 (this does not include He). The inverse of MGDR is
the molecular gas depletion time, $\tau_{\mathrm {M}}$, which represents
the time it takes to consume all of the molecular gas at the current rate
of star formation.

Recently, \cite{Leroy2008} have measured MGDR($z$ = 0) from a comprehensive
analysis of 23 nearby galaxies. The results are based on \hi\
  surface densities measured from the THINGS \hi\ survey
  \citep{Walter2008}, H$_2$ surface densities inferred from the BIMA SONG
  \citep{Helfer2003} and HERACLES (Leroy et al. 2008b) CO surveys, and star
  formation rates from both SINGS \citep{Kennicutt2003} and GALEX
  \citep{GildePaz2007} data.  The galaxies surveyed include spiral and
dwarf galaxies and the analysis was done on a pixel-by-pixel basis
convolved to a common resolution, typically $\sim$800 pc. This comparison
is the most extensive work done to date and the authors find a remarkable
constancy of MGDR over a wide range of conditions: 
$0.525 \pm 0.25 {\rm\ Gyr}^{-1}$, equivalent to a molecular gas
depletion time of 1.9 Gyr.  Their measured star formation rates and \h2
column densities vary by three orders of magnitude averaged over entire
galaxies, and the pixel-by-pixel values vary even more.  Thus the constancy
of the MGDR occurs over a wide range of conditions both within galaxies
(including nuclei and disks) and from galaxy to galaxy.

\subsubsection{Measurements at high z}
\label{MGDRhighz}
The \cite{Leroy2008} study only applies to galaxies
near $z=0$.  To extend the range of $z$ we appeal to
observations of the MGDR in normal, $z \ga 1$ galaxies.
\cite{Daddi2009} report the SFR and total gas mass for six,
  near-infrared selected BzK galaxies at $z\sim 1.5$.  Using numerical
  simulations, they calculate a conversion factor $\alpha_{CO} = M_{gas} /
  L_{CO} = 3.6 \pm 0.8$ M$_{\odot}$ (K km s$^{-1}$ pc$^2$)$^{-1}$.  This
  value is close to a Milky Way-like value of $\alpha_{CO} \sim 4.6$ and
  excludes a typical ULIRG value of $\alpha_{CO} \sim1$.  This gas mass
  includes He, so we divide by $1.4$ to calculate the MGDR, which we have
  defined to not include He. The MGDR values for these six galaxies in
  \cite{Daddi2009} vary between 1.9 and 4.7 Gyr$^{-1}$, three to nine times
  the \cite{Leroy2008} value for $z=0$. Apparently, molecular gas is 
  consumed by star formation at a much higher rate at high redshift than it is today.

  Similarly, \cite{t2010} have made an extensive survey of normal, star
  forming galaxies at redshifts 1 and 2, measuring the MGDR for a
  sample of 19 galaxies.  At each redshift locus, the selected galaxies
  sample the high mass end of the main sequence galaxy population in the
  M$_{*}$-SFR plane. The \cite{t2010} results use a Milky Way-like value
  for $\alpha_{CO}$.  They find MGDR values in agreement with the
  \cite{Daddi2009} values at $z\sim1.5$.  The data points from these
  studies are plotted in Figures~\ref{CBSFEevol} and \ref{OBfig}:
  \cite{Daddi2009} as purple diamonds and \cite{t2010} as green squares.

  In samples of galaxies selected by different optical and
  near-IR criteria, \cite{Reddy2005} find that BzK, BX/BM and DRG selected
  galaxies account for an SFRD of $\sim0.1$ M$_\sun \permpccu$ yr$^{-1}$ in
  the range $1.4 < z < 2.6$. This is most of the observed SFRD (see
  Fig. \ref{SFRfig}).  More recently, \cite{Reddy2008} find that galaxies
  with $L_{\mathrm{IR}\,}\approx L_{\mathrm{bol}\,}\lesssim 10^{12}\
  L_{\odot } $ account for $\approx70\%$ of the SFRD at $1.9\leq z< 2.7$.
  The BX/BM, BzK and DRG selection criteria sample luminous, star forming
  galaxies with L $\sim 10^{11}-10^{12}$ L$_{\odot}$ and SFR $\sim 10-500$
  M$_\sun$ yr$^{-1}$ \citep{Tacconi2008}.  Therefore, these systems sampled
  by \cite{Daddi2009} and \cite{t2010} account for most of the SFRD at
  these redshifts so that the MGDR value that typifies these galaxies
  should describe the average cosmic MGDR in Eq. \ref{sfe}. This motivates
  us to use these samples to constrain our guessed forms of the MGDR at
  other redshifts.  

Numerous other authors have estimated the MGDR from individual high
redshift galaxies, or from starbursts or ULIRGS (e.g. \citealt{gs2004}),
and all have found that the MGDR is greater in these galaxies than is
typical of galaxies at $z$ = 0.  In fact, there is no observational
evidence for a declining MGDR with increasing redshift to at least $z\sim
4$.  Taken together, all of these observations lead us to reject any model
of gas consumption that {\it requires} lower MGDR values at redshifts
significantly greater than zero.

\subsection{\rh2}
\label{h2obs}

Using a combination of CO and \hi\ measurements in the local universe,
\cite{OR2009} have determined the density of \h2 at $z = 0$, $\mrh2(0)$, to
be 1.9 -- 2.8 $\times 10^7$ h M$_\sun\permpccu$. 
The range of values corresponds to different assumptions
about how metallicity affects the CO-to-\h2~conversion ratio (the 
authors quote an error of $\pm 40\%$ on each individual calculated value). Using $h=0.7$
gives $\mrh2(0) = (1.3 - 2.0) \times 10^7$ M$_\sun\permpccu$, or
an average value of $1.65 \times 10^7$ M$_\sun\permpccu$. This value is
within 50\% of the estimate of $\mrh2(0) = 1.1 \times 10^7$
M$_\sun\permpccu$ 
by \cite{zp2006} (no error quoted) using a
different set of observations and should therefore be reasonably reliable.

\begin{figure}[ht]
\includegraphics[scale=0.5]{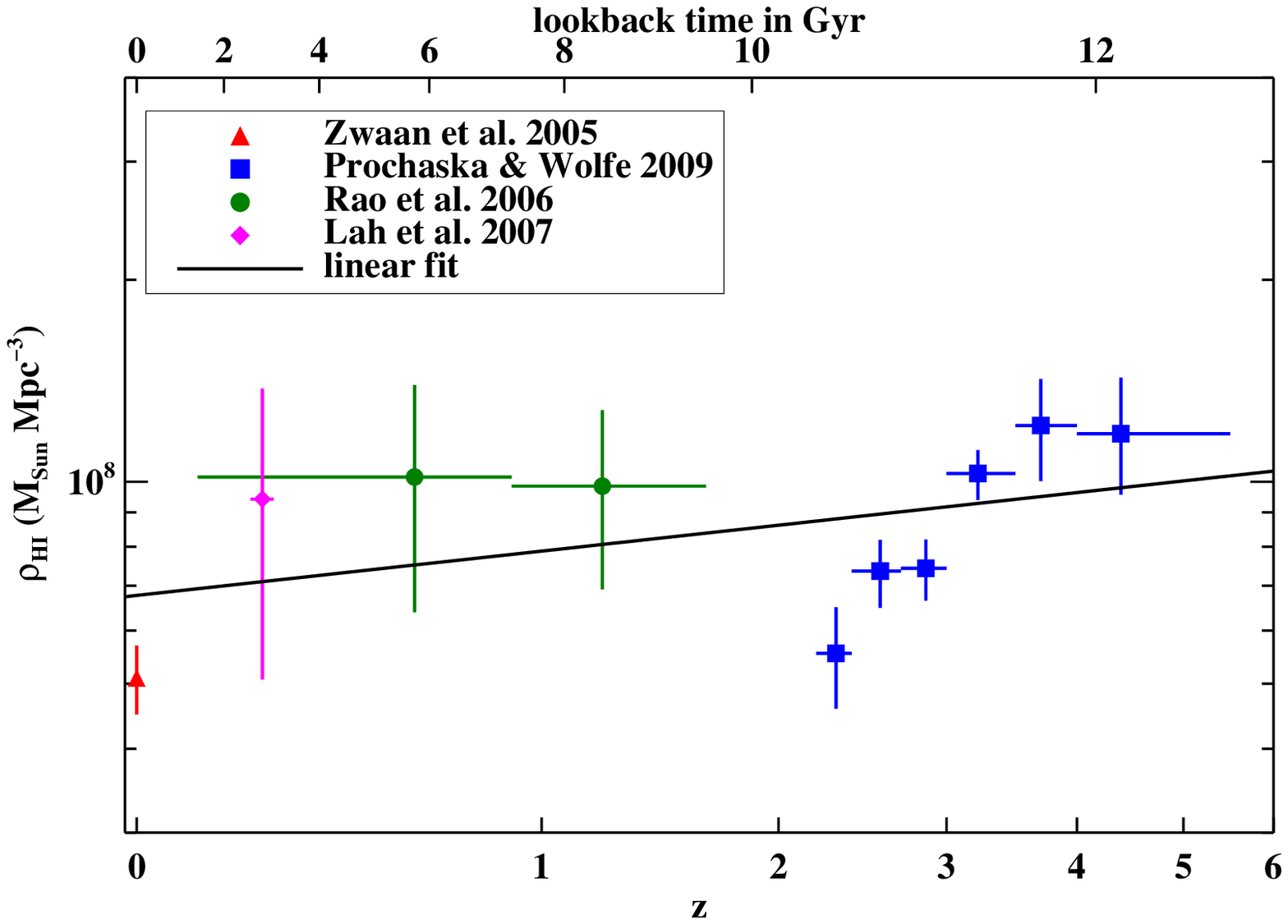}
\caption{Data points for the comoving \rhi from
  \cite{Zwaan2005,Rao2006,Lah2007};  and \cite{PW2009} with linear fit in
  $\log(1+z)$ vs $\log(\mrhi)$ space. }
\label{HIfig}
\end{figure}

\subsection{\rhi}
\label{HIobs}

A number of recent studies have reported measurements of the mean \hi\
comoving mass density in galaxies, $\mrhi(z)$, from the local universe to
$z\sim 6$.  At $z=0$, \cite{Zwaan2005} present a study of the \hi\ mass
function using the \hi\ Parkes All Sky Survey (HIPASS) data.  At low
redshift, \cite{Lah2007} calculate \rhi by co-adding \hi\ 21-cm emission
from galaxies with known positions and redshifts.  At higher redshifts,
estimates of \rhi are mainly obtained from DLA studies \citep{Rao2006,
  PW2009}.  We use these observations of \hi\ to construct an analytical
form for \rhi as a function of $z$.  We fit a straight line to these points
in $\log(1+z)$ vs $\log(\mrhi)$ space. The data points and the fit are
shown in Fig.~\ref{HIfig}. The observations suggest very little evolution
of \rhi, which increases by only a factor of 2-3 between $z=0$ and $z=6$.
A more recent DLA study \citep{Noterdaeme2009} using SDSS DR7
  finds the values at $z\sim 2$ to 3 to be somewhat higher than the
  \cite{PW2009} points; this would make our linear fit in
  Fig.~\ref{HIfig} correspond even better to the observations.
 

\section{Building a Model to Fit the Observations}
\label{models}

The observations above allow us to compare the evolution of the SFRD
and $d\mrh2/dt$.  We start with the simplest model
possible: a closed box of \h2 being turned into stars at the rate of
$d\mrh2/dt$.  In this closed box model, we initially assume
that the MGDR
is constant in time (the restricted closed box model; \S\ref{RCBM}) and
subsequently relax this condition (the general closed box model; \S\ref{closedbox}).
The failure of both models motivates us to consider an open box model
(\S\ref{openbox}), where we allow the densities of
all four phases of the IGM -- stars, \h2,
\hi\, and \hii\ -- to vary to match the observational constraints.

We assume that the molecular gas is depleted only through star formation,
and that any $\mathrm{H_2}$ dissociated or ionized by star formation
is instantaneously returned to the molecular state.  Given the short
timescales for the formation of molecular gas from its atomic form,
$\sim10^6$ yr at the relevant densities \citep{hs1971,ct2004},
this approximation should be a good one for the purposes of this
paper.  In any event, if we define $d\rho_{H_2}/dt$ to be the {\it
net} flow rate of molecular gas into stars, then there is no
ambiguity.

We write the statement that star formation
occurs only through the depletion of molecular gas:
\begin{equation}
\label{sfe}
SFRD = MGDR \times \rho_{H_2} \,.
\end{equation}
This equation was used to infer the individual MGDR 
  for a sample of nearby galaxies with
  a wide of range of SFRs and $H_2$ column densities, where it was found
that $MGDR \approx 0.5$ Gyr$^{-1}$ to a remarkable constancy
(\citealt{Leroy2008}, see \S\ref{mgdrsec}).  We use this prescription for
the SFRD for our models but we note that the form may change at higher
redshift (See \S\ref{SFRDvar}).

It is also interesting to note that if we divide the observed global star
  formation rate density SFRD$(z=0) \sim$ (0.8 -- 1.8) $\times 10^{-2}
  M_\sun\permpccu$ yr$^{-1}$ \citep{HB2006, Salim} by the observed
  $\mrh2(0) = (1.3 - 2.0) \times 10^7 M_\sun\permpccu$ \citep{OR2009}, we
  obtain a range of MGDR that is consistent with the values of
  \cite{Leroy2008}.
Given that stars must form from
molecular gas, this result is not surprising.  Nevertheless, the agreement
is reassuring because it is based on different data sets and different
methods of determining the relevant quantities.  It also suggests, combined
with the arguments in \S\ref{intro} that Eq. \ref{sfe} can be
extrapolated to all $z$.

We use mass densities instead of mass surface densities, which are used by
observers, but note that these are roughly equivalent because for the most
part, the stars, \hi\ and \h2 are generally confined to thin disks within
galaxies.

\subsection {The Restricted Closed Box Model}
\label{RCBM}

In the closed box model, we consider only stars and \h2, and allow
$\rho_{H_2}$ to be converted into stars at the star formation rate density,
SFRD:
\begin{equation}
  \frac{d\mrh2}{dt} = -SFRD \,.
\label{CBder}
\end{equation}


For the moment, we consider a restricted closed box model in which we
  take the MGDR to be constant as a function of redshift. 
  To assess the ability of this model to fit observations,
  we first combine eqs.~(1) and (2) to obtain $d(SFRD)/dt = MGDR\times
  SFRD$.  We then note that our piecewise linear fit to the observed SFRD
  as a function of time (Appendix A) implies $d(SFRD)/dt \sim (.24\, {\rm
    Gyr}^{-1}) \times SFRD$, where the coefficient is about half of the
  observed MGDR \citep{Leroy2008}.

  In other words, from the assumption of a
  constant MGDR at the present epoch in a closed box H$_2$ model, we find
  that the star formation rate is declining only half as fast as expected
  given our current reservoir of molecular gas.
It is this discrepancy in the derivatives of the observed cosmic star
formation rate and the rate at which we observe molecular gas being
converted into stars that we call the cosmic molecular gas depletion
problem.  We note here that given the uncertainties in the observations,
this factor of two in itself is not a strong argument against the closed 
box model, but we will show in general that observational constraints rule out {\it any} closed
box model.

\subsection{The General Closed Box Model} 
\label{closedbox}

\begin{figure}[htp]
\includegraphics[scale=0.55]{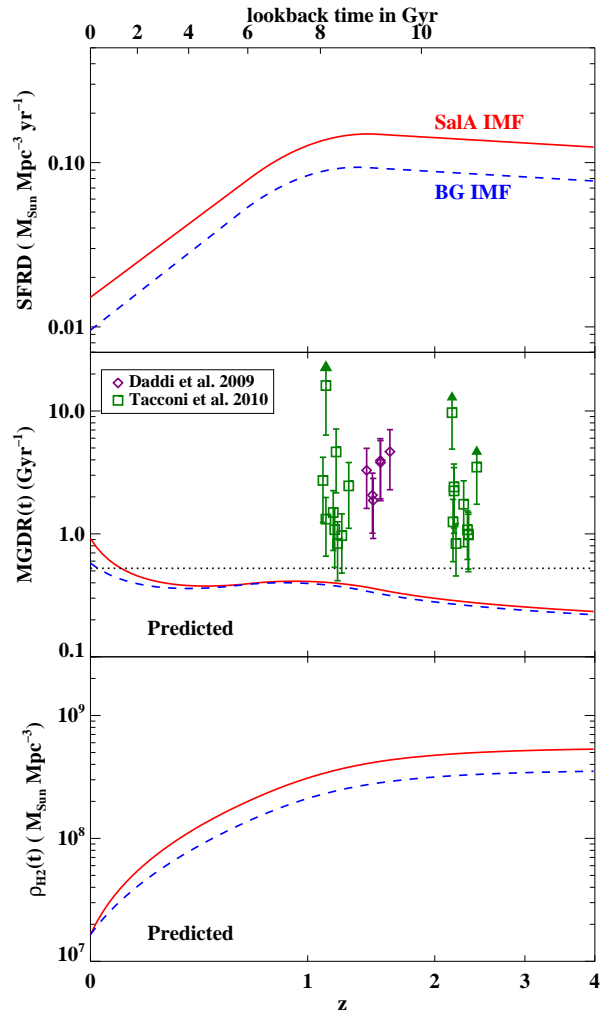}
\caption{The predicted MGDR(t) and $\mrh2(t)$ in the closed box 
model. The top panel shows the input SFRD forms (the smoothed 
piecewise linear fits from Fig. \ref{SFRfig}). The middle and
bottom panels show the MGDR and $\mrh2(t)$, 
respectively, predicted
by the general closed box model for these observed SFRD$(t)$ forms.
The points in the middle panel show the MGDR observations from
\cite{Daddi2009} (purple diamonds) and from \cite{t2010} (green squares), 
discussed in \S \ref{mgdrsec}. This model 
requires a lower MGDR in the past, contrary to observations. }
\label{CBSFEevol}
\end{figure}

We now relax the assumption of a constant MGDR in the closed box model in
\S\ref{RCBM}.  To study the predictions of this model, we calculate $\mrh2(z)$
by integrating Eq.~(\ref{CBder}) and using the observed SFRD$(z)$ as an
input.  We then divide SFRD$(z)$ by $\mrh2(z)$ to obtain MGDR$(z)$.  The
results are shown in Fig.~\ref{CBSFEevol}, where $\mrh2(0)$ is set to the
mean value from \cite{OR2009}.  The uncertainties in the SFRD (due to the IMF)
discussed in \S\ref{SFRDobs} are seen to have only a minor effect on the
resulting $\mrh2(z)$ and MGDR$(z)$.

Fig.~\ref{CBSFEevol} shows that $\mrh2(t)$ increases by a
factor of $\sim 10$ from $z=0$ to 1, and MGDR decreases with increasing
redshift, {\it contrary to the observational results discussed in \S\ref{mgdrsec}}.  Thus,
even the general closed box model is at odds with the observations, leading
us to our next model.

\subsection{The Open Box Model}
\label{openbox}


Since a closed box model of only \h2 and stars is inconsistent with
observations, we now allow additional components that can be converted into
H$_2$ and then into stars.  We consider separately the \hi\ gas and an
external source of gas that we call $\mrext$, and modify Eq.~(\ref{CBder})
to
\begin{equation}
  \frac{d\mrh2}{dt} = -SFRD - \frac{d\mrhi}{dt} - \frac{d\mrext}{dt}  \,.
\label{WOBder2}
\end{equation}
\subsubsection {\rhi}
\begin{figure}[ht]
\includegraphics[scale=0.44]{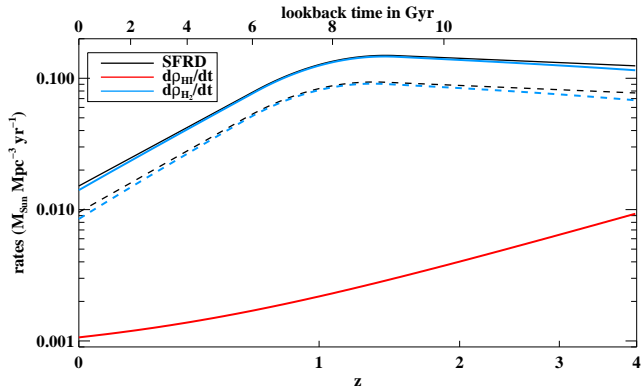}
\caption{ Rates of gas flowing from one phase to another
in an open box model with only stars, \h2 and HI (ignoring \rext). 
The SFRD (black curves) is an order of magnitude higher than $d\mrhi/dt$
(red curve), forcing $d\mrh2/dt$ (blue curves)
to be approximately equal to the SFRD, as in the closed box model. 
The two forms of the SFRD described in \S \ref{SFRDobs} are used:
modified Sal A IMF (solid) and the \cite{BG2003} IMF (dashed).
The derivatives
of \rhi and \rh2 are negative, but the absolute values are plotted here.}
\label{HIder}
\end{figure}

For the \hi\ gas, the observations discussed in \S\ref{HIobs} and Fig.~\ref{HIfig}
suggest that $\mrhi(z)$ is very slowly varying over cosmic timescales;
therefore $d\mrhi/dt$ is small. Fig. \ref{HIder} shows that the derivative
of the observed \rhi (red curve) is an order of magnitude smaller than the observed
SFRD (black curves).  In the absence of $\mrext$, we have
$|d\mrh2/dt|$ (blue curves) is approximately equal to $SFRD$, as in the failed
closed box model. Thus the inclusion of \hi\ {\it alone} in an open-box model
is not enough to fit the model to the observations.  This leads to a
robust conclusion: {\it the reservoirs of \hi\ and \h2 at all times in the
past (at least as far back as $z=4$) are insufficient to fuel the star
formation over cosmic timescales.}

It is important at this point to clarify that \rhi represents the reservoir
of \hi\ both in galaxies as well as the \hi\ outside of galaxies.  This is
because the DLA observations of Prochaska and Wolfe (2009) include all of
the high column density neutral \hi\ (N(HI) $> 2 \times$ 10$^{20}$ \c2).
This gas contains at least 85\% of the neutral \hi\ atoms for $z < 6$
(O'Meara et al 2007).

\subsubsection {$\rho_{ext}$}

We are therefore forced to include a nonzero $d\mrext/dt$ term in the open
box model.  This component represents the ionized intergalactic gas at all
temperatures and densities that can recombine to form \hi\ within a Hubble
time.  Effectively, it is the ionized gas in the filaments of the cosmic
web.  

Note that some of the \hi\ can become ionized and redistributed to
the intergalactic medium at high enough temperatures and low densities such
that this gas does not recombine in a Hubble time.  Such gas can be ejected
by means of supernovae, galactic winds or AGN. However, if we define
$d\rho_{ext}/dt$ to be the {\it net} flow out of the ionized phase, then the
gas that is fed back into the ionized state is implicitly included in our
accounting.  
That is, any \hi\ fed back into the ionized phase is made up by
an equivalent increase in $d\rho_{ext}/dt$. If the total amount of ionized gas
available is represented by $\Omega_{baryon}$ minus the total amount
of baryons in galaxies, we may consider the ionized phase to be 
a nearly infinite reservoir of gas
available to fuel star formation.  Any gas ionized and added to that
reservoir by star-formation and active galaxies is negligible. 
 


\begin{figure*}[htp]
\includegraphics[scale=0.55]{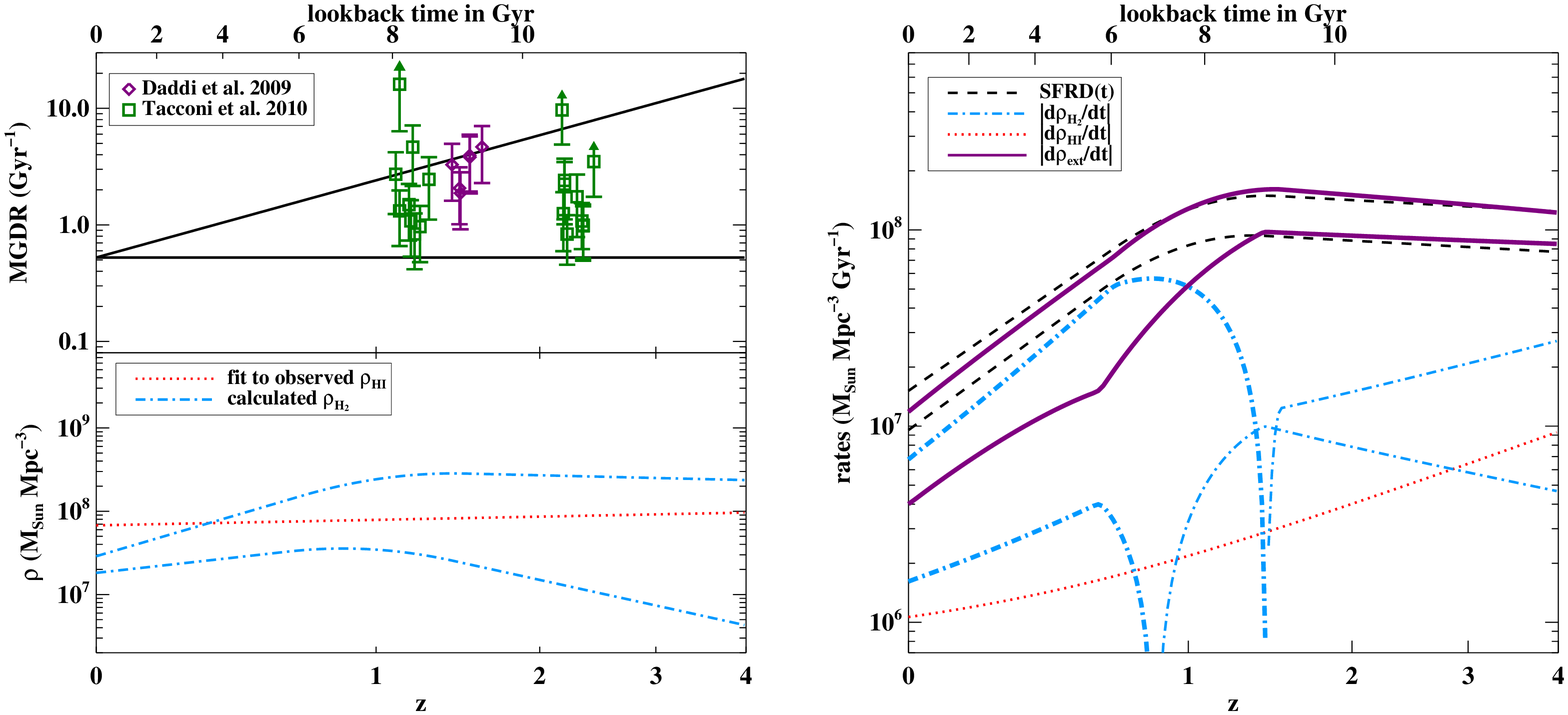}
\caption{Predictions of the open box model, calculated from combinations of
  the inputs: two forms of the MGDR (solid black lines in top left panel), 
  two forms of the SFRD (dashed black lines in right panel) and
  the fit to the observations of \rhi (red dotted line in bottom left panel).  In the top left panel,
  we plot the MGDR points from \cite{t2010} (green squares) and
  \cite{Daddi2009} (purple diamonds) to
  constrain the guessed range of MGDR forms.  The possible range of each of
  the calculated quantities, $d\mrh2/dt$, $d\mrext/dt$ and \rh2, is shown
  by two lines illustrating the minimum and maximum curves from the four
  combinations of inputs. In the right panel, the absolute values of the rates are plotted:
  thicker lines indicate negative values. }
\label{OBfig}
\end{figure*}

To compute the $d\rho_{ext}/dt(t)$ required in the open box model to
match the observations in \S 2, we start with an observed SFRD$(t)$ and
$\mrhi(t)$, and a guess for the form of the MGDR$(t)$ that
is compatible with the data points from \cite{t2010} and
\cite{Daddi2009}.  From these inputs, we compute $\mrh2(t)$ and its time
derivative using $\mrh2(t) = \mathrm{SFRD}(t) / \mathrm{MGDR}(t)$.
Combining $d\mrh2/dt$ with the observed SFRD$(t)$ and $d\rho_{HI}(t)/dt$ in
Eq.~(\ref{WOBder2}) then gives $d\mrext/dt$.
The results of this procedure are shown in Fig. \ref{OBfig}.  

To illustrate the expected range of possible values, we use two forms for
the input MGDR (top left panel) and two forms for the input SFRD (black dashed curves in
the right panel).  We bracket the possible values for the MGDR on one side as
constant using the measured value at $z=0$, and on the other as one that
increases linearly to larger redshifts as suggested by the \cite{t2010} and
\cite{Daddi2009} data points.  For the SFRD, we use the smoothed piecewise
fits from \cite{HB2006} for the two extreme IMFs discussed in \S\ref{SFRDobs}. 
We then calculate $d\mrh2/dt$, $d\mrext/dt$ and \rh2 for each of the four
possible combinations of MGDR and SFRD.  The range of calculated values is
indicated in Fig. \ref{OBfig} by plotting the minimum and maximum of the four curves for each
calculated quantity.

In the bottom left panel, we note that the minimum \rh2 curve lies below \rhi at
all times, and the maximum curve becomes larger than \rhi at $z \sim
0.35$. The difference between the two curves is mostly the MGDR choice: the
maximum curve corresponds to the higher SFRD and the flat MGDR, while the
minimum curve corresponds to the lower SFR and the increasing MGDR (as
expected from Eq. \ref{sfe}).  All other combinations of MGDR and SFRD
whould lie between these two curves. {\it Because all of the molecular
gas resides in galaxies, we require that the molecular gas mass in galaxies
is larger, on average, at $z$ = 1-2, than is typical at $z$ = 0.}  The
change in \rh2 resulting from changing the SFRD is minor. Therefore,
observations of \rh2 at $z > 0$ or the redshift at which $\mrh2 = \mrhi$
will constrain the form of the MGDR.

In the right panel, the absolute value of the rates is plotted, with
a thicker line style used to indicate negative rates. The negative
portions of the $d\mrh2/dt$ curves correspond to decreasing \rh2 as we
move forward in time towards $z=0$, as \h2 is converted into
stars. The $d\mrext/dt$ curves are negative for the whole range of
redshifts plotted, indicating a flow of external gas into the \hi\
reservoir of galaxies.  We note that the range of solutions for
$d\mrext/dt$ does not deviate very much from the SFRD: roughly
a factor of two at low redshifts for the minimum case.  This is
because the reservoirs of H$_2$ and \hi\ are so small compared to what is
required by the observed SFRD. {\it Therefore, we
conclude that the amount of inflow needed from this external gas,
$d\mrext/dt$, is approximately equal to the SFRD.} 
This observational conclusion is reinforced by cosmological simulations 
which find that star formation rates closely follow gas infall rates \citep{Keres2005,Dekel2009}.
Specifically, \cite{Keres2009} calculate an upper limit on the 
external gas supply feeding galaxies that is only a factor of 2 higher
than our predicted $d\mrext/dt$ curves and shows similar evolution with redshift.


\section{Discussion}
\label{discussion}

We have shown in \S\ref{RCBM} and \S\ref{closedbox} why the closed \h2 box model
doesn't work.  Here we discuss the open box model of \S\ref{openbox} and what this model predicts.

\subsection{Variations in the SFRD} 
\label{SFRDvar}

For the models in \S\ref{models}, we have extended the star formation rate prescription from
\cite{Leroy2008} to mass densities averaged over Mpc scales: SFRD $\propto$ \rh2. 
Some studies of galaxies
with higher gas surface densities have found evidence for a steeper power law of
SFRD $\propto \mrh2^{1.4}$ \citep{k98, wong2002, Bouche2007}.  The
choice of SFRD prescription at a given redshift depends on what type of
galaxies dominate the SFRD at that redshift. At the present epoch, regular
spiral galaxies, such as those studied by \cite{Leroy2008}, seem to
dominate the SFRD: \cite{Salim} finds that galaxies in the mass range
$10^{9.3} < M_* < 10^{10.6} M_\odot$ account for about half of the total
SFRD.  At higher redshifts, however, the BzK and BX/BM galaxies dominate
the SFRD (see \S \ref{MGDRhighz}). These galaxies are more gas rich, and thus
the $\mrh2^{1.4}$ prescription may be more appropriate.
 
For this paper, we have assumed the \cite{Leroy2008} prescription for the
SFRD, but note that the power law may change at higher redshift because the
dominant mode of star formation may change.  This is equivalent to a change
in the MGDR with time, a case we consider explicitly and probably
contributes to the values of the MGDR found by \cite{t2010} and
\cite{Daddi2009}.  If, for example, $SFRD = MGDR(0) \mrh2^{1.4}$ at
higher redshift, then we would find $MGDR(z) = MGDR(0) \mrh2^{0.4}$ by
forcing our prescription of $SFRD = MGDR \times \mrh2$.  Thus as \rh2
increases at higher redshift, an increase in the power law of the SFRD
prescription will be manifested as an increase in the MGDR. An increase
would tend to bring the MGDR closer to the upper bound in the top panel of
Fig.~\ref{OBfig}, with the result that $\rho_{H_2}(z)$ would lie closer to
the lower blue curve in the bottom panel in Fig.~\ref{OBfig}.

\subsection{Stellar Recycling}
Stellar recycling is an important component of any treatment of gas 
evolution in galaxies. \cite{k94}, for example,
suggest that gas returned to the ISM during stellar evolution can
signficantly increase the gas depletion time in the Milky Way and in
nearby galaxies. In the model presented here, we do not explicitly consider the effect
of gas return to the ISM in Eqs. \ref{sfe} and \ref{WOBder2},
but we argue that our formulation already accounts for
stellar recycling due to the nature of the observed quantities we
use as inputs. Most of the return of gas to the ISM comes from red giant
stars (see e.g. \citealt{Blitz1997}), and much of that is returned to the ISM
in the form of molecules \citep{Marengo2009}. Even
gas that is returned in other phases becomes largely
molecular after each spiral arm passage, the timescale for which is
$\sim 10^8$ yr in most galaxies, a timescle short compared to the
rather long timescales we consider in this paper. Since the recycled
gas is largely molecular, and any that isn't is quickly returned to the
molecular phase, we argue that observations of molecular gas
already include the 
gas returned via stellar recycling. Therefore, this recycled component 
is included in our initial condition, $\mrh2(z=0)$, as well as the MGDR.
Since we use the MGDR to relate SFRD and \rh2 at each time step, 
stellar recycling is built into our model through these observations, so
we need not include an explicit recycling term in our equations.

\subsection{Behavior of $\mrh2(z)$}

Although the exact shape of \rh2 depends sensitively on the form of MGDR,
we have bounded the behavior of \rh2 by calculating what we take to be
limiting cases in
our open box model (Fig.~\ref{OBfig}).  The \rh2 curves all rise with
increasing redshift, peaking around $z \sim 1 - 1.5$ at 1.5 to 10 times the
value of \rh2 today.  After the peak, \rh2 may fall off toward higher
redshift if the MGDR is rather constant, or stay close to constant if the
MGDR remains high.  The prediction is less well constrained at higher
redshifts. It is noteworthy that \cite{t2010} find that the gas disks they
observe at $z=1$ and $z=2$ have considerably more molecular gas relative to
the stars, typically about 30-50\%, compared to $\sim 1- 5\%$ for the Milky
Way and nearby disk galaxies \citep{Helfer2003}.  This trend is consistent
with our estimates.  Although there is some uncertainty in the \h2
masses of \cite{t2010} because of the uncertainty in the value of X$_{CO}$,
there appears to be little doubt that the ratio of \h2 to stellar mass in
the BzK and BX/BM galaxies is higher than typical values for similar
galaxies at $z=0$.  Due to the sensitivity of \rh2 to the form of MGDR,
future observations of \rh2 or the ratio \rh2/\rhi at higher
redshift would allow us to better constrain the evolution of MGDR, and
reduce the area between the bounding curves of Fig.~\ref{OBfig}.

\subsection{The Nature of $d\mrext/dt$}
\label{rextnature}

In our open box model, the \hi\ reservoir, $\mrhi$, is augmented by an inflow of
gas from \rext at a rate of $10^7 - 10^8 M_\sun \permpccu$ Gyr$^{-1}$,
depending on $z$.  This high rate of inflow means that the gas being
accreted is mostly ionized since the fraction of neutral hydrogen outside
of the \rhi reservoir is too small.

The \rhi inferred from observations of DLA systems accounts for the \hi\
associated with galactic disks. As mentioned in \S\ref{openbox}, \cite{OMeara2007}
find that systems with column density $\Sigma_{HI} < 2 \times 10^{20}$
cm$^{-2}$ account for $\approx 15\%$ of neutral hydrogen atoms at all $z < 6$.
Therefore, the fraction of \hi\ outside DLA
systems is about $15\%$, corresponding to roughly $1.5 \times 10^7$ M$_\sun
\permpccu$.  For an average inflow rate of a few times $10^7$ M$_\sun
\permpccu$ Gyr$^{-1}$ for the past 10 Gyr, this intergalactic \hi\ could only
account for ten percent of the total, at most.  
Therefore, the inflow of gas needed for fueling
ongoing star formation represented by $d\mrext/dt$ must be almost
completely ionized.

Recently, cold flows have been suggested as an important
source of gas for galaxy formation and evolution 
\citep{Keres2005,dekel06}.  In these models,
galactic disks in halos with $M \la 10^{12} M_\sun$ are built up by direct
accretion of cold gaseous streams from the cosmic web.  For galaxies with
larger masses, cold flows are also the dominant means of mass accretion,
but different outcomes for individual galaxies depend on the epoch of
inflow.  If this picture is correct, our work implies that the cold flows
must be almost entirely ionized.

The same is true if the gas needed to fuel star formation is brought in
primarily through minor mergers.  If this gas were in atomic form, it would
be part of the inventory of atomic gas observed in the DLA systems, which
we have shown in \S\ref{openbox} to contribute negligibly to fueling the
star formation at all redshifts up to $z = 4$.

\subsection{Cooling Times}

The open box model
requires $d\mrext/dt \sim SFRD$, or about $10^7$ to $10^8$ M$_\sun
\permpccu$ Gy$^{-1}$.  We use these numbers for $d\mrext/dt$ to calculate a
cooling time for the gas in the context of two models for gas accretion
onto galaxies: two-phase cooling of hot halo gas \citep{MB2004} and cold
flow accretion \citep{KH2009}.

We estimate the cooling time, $t_{cool}$, by taking
\begin{equation}
\frac{\rho_{gas}}{t_{cool}} \sim \dot{\rho}_{ext} \,,
\label{cool1}
\end{equation}
where $\rho_{gas}$ is the average mass density of the cooling ionized gas
smoothed over the appropriate volume (to be specified for each
cooling model individually). We set $\rho_{gas}$ equal to $m_{H} n_e f$
where $n_e$ is the local number density of electrons and $f$ is the filling
factor for the relevant volumes ($\bar{n}_e/n_e$).  Combining this with
the cooling time
\begin{equation}
t_{cool} \sim \frac{3 k_b T}{2 \Lambda(T) n_e} \,,
\label{cooltime}
\end{equation}
where $\Lambda(T)$ is the cooling function of the gas, gives
\begin{equation}
f n_e^2 \sim \frac{3 k_b T \dot{\rho}_{ext}}{2 \Lambda(T) m_H}  \,.
\label{cool2}
\end{equation}
As a basis for comparison, we first estimate the filling factor $f$ for hot halos
of $L_*$ galaxies out to the virial radius.
We make the simplistic assumption that the universe is made up of $L_*$
galaxies with masses $M_{dyn} \sim 10^{12}$ M$_\sun$ and 
circular velocities of $\sim160$ km s$^{-1}$. Therefore, the 
virial radius, $R_{vir} = GM_{dyn}/v^2 \sim 300$ kpc. 
We estimate the average number density in this simple
universe composed of $L_*$ galaxies by dividing the total luminosity
density, $\mathcal{L}$, by $L_*$. We use the r band values from \cite{Blanton2003}
which calculates the galaxy luminosity function at $z\sim0.1$ from SDSS data: 
$\mathcal{L} \approx 1.84 \times 10^8$ h M$_{\odot}$ Mpc$^{-3}$, 
$L_* \approx  1.2 \times 10^{10}$ h$^{-2}$ M$_{\odot}$. This yields
$n_{L_*} \sim \mathcal{L} / L_* \sim 0.015$ h$^3$ Mpc$^{-3} \sim 5 \times 10^{-3}$ Mpc$^{-3}$.
More recent work on the luminosity function using SDSS DR6 yields similar
results \citep{MD2009}. Therefore, in this simple universe, the filling factor for the $L_*$
galaxy halos is
\begin{equation}
f \sim n_{L_*} \frac{4}{3} \pi R_{vir}^3 \sim 6 \times 10^{-4} \,.
\label{cool4}
\end{equation}

\cite{MB2004} consider gas within the cooling radius of a halo, $R_c$, 
cooling via cloud fragmentation. This results in the formation
of warm ($\sim 10^4$ K) clouds within the hot gas halo. In this model for
the two-phase cooling of the hot halo gas, the relevant temperature for 
the gas is the virial temperature of the halo, $\sim 10^6$ K for a Milky Way type
halo.



\cite{KH2009} find that the majority of cold clouds that form 
around Milky Way type galaxies are the result of filamentary 
"cold mode" accretion. Most of the gas does not reach the virial
temperature of the halo, $\sim 10^6$ K, but rather cools from
a maximum temperature of $\sim 10^4 - 10^5$ K. 


\begin{figure}[ht]
\includegraphics[scale=0.5]{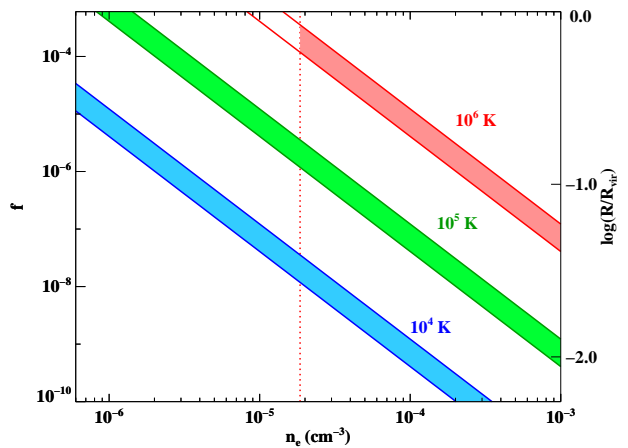}
\caption{ 
Shaded regions show the allowed filling fraction $f$ and $n_e$ for three temperatures:
$10^4$ K in blue, $10^5$ K in green and $10^6$ K in red. 
The axis on the right gives the radius of the relevant volumes as a fraction of the virial
radius corresponding to the filling fraction $f$ in  
a simple universe filled with $L_*$ galaxies.
At each temperature, the allowed region of $n_e$ vs $f$ space is 
calculated using the predicted range for the inflow rate at
$z = 0$: $d\mrext/dt = 0.4 - 1.2 \times 10^7$ M$_\sun$ $\permpccu$ Gyr$^{-1}$.
The region is further bounded by the vertical dotted line
at the value of $n_e$ corresponding to 
a cooling time equal to the age of the universe. 
This line is only plotted for the $10^6$ K gas: for
the other two temperatures, the densities are very small
and lie outside of the plotting range.
}
\label{density}
\end{figure}

We examine the possible values for $n_e$ and $f$ in these two models by
calculating $f$ as a function of $n_e$ for three temperatures: $10^4$ K and $10^5$ K
for cold flow accretion and $10^6$ K for cooling from the hot halo. In
Fig. \ref{density}, we plot $f$ versus $n_e$ for these three temperatues
for the range of $d\mrext/dt$ at $z=0$ predicted by our open box
model: $0.4 - 1.2 \times 10^7$ M$_\sun$ $\permpccu$ Gyr$^{-1}$. The
y-axis on the right side shows $\log(R/R_{virial})$ corresponding to
$f$, where R is the radius of the relevant volume associated with an $L_*$
galaxy. The vertical dotted line indicates the value of $n_e$ at which the
cooling time is equal to the age of the universe.  We use the
approximate form for $\Lambda(T)$ for mildly enriched gas ($Z = 0.1$)
from \cite{MB2004} :

\begin{equation}
\Lambda(T) \simeq 2.6 \times 10^{-23} \left(\frac{T}{10^6\ K}\right)^{-1} \ cm^{-3}\ erg\ s^{-1}
\label{lambdaeq}
\end{equation}

For the cold flow gas at $10^4$ K and $10^5$ K, the dotted lines where the cooling times
equal the age of the universe are outside of the plotting range, so the 
minimum allowed value of $n_e$ is where $R \sim R_{virial}$: about $8 \times 10^{-7}$ cm$^{-3}$
for the $10^5$ K gas and even smaller for the $10^4$ K gas.
For the cooling hot halo gas at $10^6$ K, 
the condition that the cooling time be less than the age of the universe
forces $n_e$ to be larger than about $2 \times 10^{-5}$ cm$^{-3}$. 

These calculations don't put strong constraints on the 
density of the halo gas since the cooling times are so rapid for a 
large range of temperatures and densities. How the gas gets into the galaxies
themselves will involve a more complete treatment including the effects 
of self-sheilding, which is beyond the scope of this paper.

\subsection{Comparing $d\mrext/dt$ to Dark Matter Accretion Rate}


The rate $d\mrext/dt$ inferred from our open box model provides an estimate
for the average rate at which the baryonic fuel is required to make its
way to a galactic disk in order to sustain the observed star formation,
which largely occurs in the disk.  A comparison of this rate with the mean
rate of baryon accretion at the virial radius of the host dark matter halo
will provide an estimate for the efficiency of converting the cosmological
infalling baryons into stars.  Many cooling and feedback processes
obviously affect the fate of baryons after their infall onto
the halo and whether they will reach the disk.  In fact,
much of the current research in galaxy formation modeling is aimed at
understanding this transition.  Our goal here is to estimate an overall
ratio, as a function of redshift, of the baryon accretion rates 
at the virial radius and at the disk scale.

We begin with the dark matter accretion rate from \cite{McBride09} and
\cite{FMBK10}, which quantified the mass accretion histories of all dark
matter halos with masses above $\sim 10^{10} $ M$_\odot $ in the two
Millennium simulations of a $\Lambda$CDM universe \citep{Springel05,BK09}.
An approximate function is provided for the average mass accretion rate as
a function of redshift and halo mass \citep{FMBK10}:
\begin{eqnarray}
        \dot{M}  &=& \beta \, M_\odot {\rm yr}^{-1} 
         \left( \frac{M}{10^{12} M_\odot} \right)^{1.1} \nonumber\\
        && \times (1 + \gamma\, z) \sqrt{\Omega_m (1+z)^3 + \Omega_\Lambda} \,. 
\label{rate}
\end{eqnarray}
where $(\beta,\gamma)=(25.3, 1.65)$ for the median rate and (46.1, 1.11) for the mean rate, 
$\Omega_m$ and $\Omega_\Lambda$ are the present-day
density parameters in matter and the cosmological constant, and $\Omega_m +
\Omega_\Lambda$ is assumed to be unity (as in the Millennium simulation).
The mean rate is $\sim 50$\% higher than the median rate due to the long tail
  of halos with high accretion rates in the distribution.
This $\dot{M}$ represents the average rate at which the mass in dark matter
is being accreted through the virial radius of a halo.  The mass growth
comes in two forms in cosmological simulations: via mergers with other
resolved halos \citep{FM09a}, and via "diffuse" accretion of non-halo
material that is a combination of unresolved halos and unbound dark matter
particles \citep{FM09b}.

We convert $\dot{M}$ above into a mean accretion rate for the baryons,
$\dot{M}_b$, by assuming a cosmic baryon-to-dark matter ratio of $f_b=
\Omega_b/\Omega_m = 1/6$.  The result should provide a reasonable
approximation for the mean rate of baryon mass that is entering the virial
radius via mergers plus accretion of intergalactic gas.  These infalling
baryons are presumably in a mixture of forms: warm-hot ionized hydrogen gas
of $10^5$ to $10^7$K, ``cold'' flows of $\sim 10^4$K (still ionized) gas,
and \hi, \h2, and stars brought in from merging galaxies.  As discussed earlier,
the majority of these baryons must be in the form of \hii\ gas.

To compare $\dot{M}_b$ with the rate of external gas inflow, 
$\dot{\rho}_{ext}$ (\S\ref{openbox}), needed to account for
the evolution of the observed star formation rates, we define 
\begin{equation}
   \alpha = \frac{ d\mrext/dt}{f_b \dot{M} M (dn/dM)}\,,
\label{alpha}
\end{equation}
where $M$ is the mass of the dark matter halo under consideration,
$\dot{M}$ is calculated using Eq.~(\ref{rate}), and $dn/dM$ is the
(comoving) number density of dark matter halos with mass in the range of
$M$ and $M+dM$.  The parameter $\alpha$ represents the fraction of
accreting baryons (at the virial radius) that must be converted into
stars in our open box models.


The value of $\alpha$ can be estimated by combining the allowed range of
$d\mrext/dt$ from Fig.~\ref{OBfig} with the halo abundance $dn/dM$ from the
Millennium simulation \citep{Springel05}.  Taking $f_b = 1/6$ and $M =
10^{12}$ M$_{\odot}$, we find the predicted $\alpha$ to be $\sim 70-100$\% at
$z\ga 3$, $\sim 120-200$\% at $z\sim 2$, and $\sim 30-90$\% at $z=0$.  We
note here that several factors may contribute to the uncertainty in the
alpha values. First, the distribution of $\dot{M}$ is broad at a
  given halo mass (see, e.g., Fig.~5 of \citealt{FMBK10}), and we have
  simply used the mean value for a rough estimate of $\alpha$ here.
Second, $\alpha$ depends on the halo mass appropriate for the population of
galaxies that dominates the SFRD at a given redshift.  Using a clustering
analysis, \cite{Adelberger} find that BzK and BX/BM galaxies have halo
masses of about $10^{12}$ M$_\sun$, but the average value appropriate to
our analysis may vary. However, since $\dot{M} \propto M^{1.1}$ and $M
dn/dm$ is approximately $\propto M^{-1}$, the mass dependence is weak, so
alpha changes by a maximum of $\sim 20$\% if we change the halo mass by a
factor of 3.  Third, since the $d\mrext/dt$ calculated in our open box
model roughly traces the SFRD, the alpha values, especially around $z \sim
1-2$, will be affected by the exact form of the SFRD.  Finally, the large
alpha value at $z\sim2$ may be reflecting a change in the fraction of
baryons in the filaments at that redshift.  Considering all the
uncertainties, we make the conservative suggestion that {\it the open box
  model requires a large fraction ($\sim 30-90$\%) of the infalling baryons
  at the virial radius to be turned into stars from $z\sim 0-4$.}  
  This is consistent with the work of \cite{Dekel2009}, which finds that
  the star formation rates in the typical 'star-forming galaxies' at
  $z\sim2$ are very close to the baryonic inflow rates from simulations.

Another way to estimate $\alpha$ is to compare the various $\dot{M}$
directly.  At $z\sim 0$, the median baryon accretion rate from eq.~(9)
  is $\dot{M}_b \sim 9$ M$_\odot$ yr$^{-1}$ for $2\times 10^{12}$ M$_\odot$
  halos, and the measured star formation rate in the Milky Way ranges from
  $\dot{M}_* \sim 2$ M$_\odot$ yr$^{-1}$ to $\sim 4$ M$_\odot$ yr$^{-1}$
  \citep{ms1979, diehl06}.  These rates imply $\alpha \sim
  \dot{M}_* / \dot{M}_b \sim 25$ to 50\% at $z\sim 0$.  In addition, we
can use the predicted $\dot{\rho}_{ext}$ from our open box models to
estimate a mean conversion rate of external \hii\ gas into stars per
galaxy, $\dot{M} \sim \dot{\rho}_{ext} / n_{L_*}$. Taking
  $\dot{\rho}_{ext} \sim 10^7$ M$_\odot {\rm Mpc}^{-3}$ Gyr$^{-1}$ from
  Fig.~\ref{OBfig} and $n_{L_*} \sim 5 \times 10^{-3}$ Mpc$^{-3}$ for the
number density of $L_*$ galaxies today, we obtain a rate of $\sim 2$
M$_\odot$ yr$^{-1}$, which is consistent with the star formation rate in
the Milky Way.


\section{Summary and Conclusions}

In this paper, we have built a quantitative model of gas consumption 
on cosmic scales based solely on observations. Using 
the observed the Star Formation Rate Density (SFRD), Molecular
Gas Depletion Rate (MGDR), and volume averaged density of molecular
hydrogen (\rh2) and atomic hydrogen ($\rho_{HI}$) 
we have defined the cosmic molecular gas depletion problem and calculated
the mass flow rates and
densities of the star-forming gas back to $z \simeq 4$
needed to resolve it.  
Extrapolations further back in time are primarily limited by uncertainties in the SFRD.
We find that:
\begin{itemize} 
\item There are no models of gas consumption where the reservoir of
gas is either
only \h2 or both \hi\ and \h2 that are compatible with the
observations of molecular gas in the galaxies at $z \sim2$ that produce the
observed SFRD.
\item Inflowing ionized intergalactic gas must provide most of the gas that
  turns into stars to times as early as $z\sim 4$. There is so little
  neutral gas inflow at all epochs, that the neutral gas ought to be
  considered more as a phase of cosmic gas flow than a reservoir of star
  forming gas.
\item 
The rate of mass inflow from the ionized state to the atomic state roughly
traces the star formation rate density. 
From $z \simeq 1 - 4$, the mass inflow rate is 1 - 2 $\times 10^8$ M$_\sun$ Mpc$^{-3}$
  Gyr$^{-1}$. At $z \la 1$, the mass inflow rate varies linearly from about
  0.5 $\times 10^7$ M$_\sun$ Mpc$^{-3}$ Gyr$^{-1}$ at $z = 0$ to about 1.5
  $\times 10^8$ M$_\sun$ Mpc$^{-3}$ Gyr$^{-1}$.  
  At all redshifts, we find the mass inflow rate 
  must be a significant fraction of the infalling rate at the virial radius, 
  in agreement with simulations. 
\item For all models, the volume averaged density of H$_2$ increases from
  its present value by a factor of 1.5 to 10 at $z$ = 1 -- 2 depending
  mostly on MGDR(t).
\end{itemize}
  
\bigskip
\bigskip
\bigskip

-------------------------------------------------------------------
\acknowledgements
This work has been partially supported by NSF grant AST-0838258.  LB
would also like to acknowledge Distinguished Visitor Awards from the
University of Sydney and CSIRO as well as a visiting scholar award
from the Center for Astrophysics, where much of the present work was
done.  We would like to thank Reinhard Genzel and Linda Tacconi for
use of their results prior to publication, and conversations with many
people including Robert Braun, Avi Loeb, Du{\v s}an Kere{\v s}, Norm Murray and
Eliot Quataert. 



\begin{appendix}

\section{Functional Form for the SFRD}
\label{appSFRD}

The form for the SFRD used throughout the paper is a smoothed
form of the piecewise fits from \cite{HB2006}. The original piecewise
function is defined by intercept and slope a and b for $z < z_1$ and c
and d for $z > z_1$ (we only consider $z<4$). The smoothed section is
just a quadratic form in the $\log(1+z)$ vs $\log(SFRD)$ plane over a
distance $2\Delta$ in $\log(1+z)$. We give the piecewise function for
the SFRD below, with $x = \log(1+z)$ and $x_1 = \log(1+z_1)$.
\begin{eqnarray} \log\left(SFRD\right) = \left\{
\begin{array}{lllr@{x}l} a &+& bx & 0 \leq & \leq x_1 - \Delta \\ a
&+& b(x_1 - \Delta) +
\frac{(d-b)}{4\Delta}\left(x^2-(x_1-\Delta)^2\right) \\ &+&
\frac{1}{2}\left[ b+d + (b-d)\frac{x_1}{\Delta}\right] \left(x - x_1 +
\Delta \right) & x_1-\Delta \leq & \leq x_1+\Delta \\ c &+& dx &
x_1+\Delta \leq & \leq x_{max} \end{array}\right.  \end{eqnarray}

In the table below, we give a,b,c,d and z1 for each of the 
two piecewise linear fits from \cite{HB2006} (for the Modified Salpeter A IMF
and the \cite{BG2003} IMF). 
To produce reasonable smoothing, we used $2\Delta = \log(1.5)$.

\begin{table}[h]
\begin{center}
\begin{tabular}{c | r @{.}l r@{.} l}
& \multicolumn{2}{c}{Mod SalA IMF} & 
\multicolumn{2}{c}{BG 2003 IMF} \\ \hline
a & -1&82 & -2&02 \\
b & 3&28 & 3&44 \\
c & -0&724 & -0&930 \\
d & -0&26 & -0&26 \\
$z_1$ & 1&04 & 0&97 \\
\end{tabular}
\caption{Table of fit parameter values for the two piecewise linear fits
from \cite{HB2006}: the Modified Salpeter A IMF and the \cite{BG2003}
IMF.}
\end{center}
\end{table}

\end{appendix}

\end{document}